\documentclass{article}
\usepackage{waspaa17,amsmath,graphicx,url,times}
\usepackage{color}
\usepackage{subcaption}
\usepackage{algorithm}
\usepackage{algorithmic}

\title{Non-uniform time-scaling of Carnatic music transients}

\newcommand{\gamaka}{{\it gamaka}}

\newcommand{\svara}{{\it svara}}
\newcommand{\Svara}{{\it Svara}}

\newcommand{\rAga}{{\it r\=aga}}
\newcommand{\RAga}{{\it R\=aga}}

\newcommand{\Gamaka}{{\it Gamaka}}

\newcommand{\varNa}{{\it var\d{n}am}}

\newcommand{\jAru}{{\it j\=aru}}
\newcommand{\kampita}{{\it kampita}}

\newcommand{\bhairavI}{{\it bhairav\=i}}
\newcommand{\kEdAragauLa}{{\it k\=ed\=aragou\d{l}a}}
\newcommand{\tODI}{{\it t\=o\d{d}\=i}}

\newcommand{\mOhana}{{\it m\=ohanam}}

\newcommand{\ZankarAbharaNa}{{\it \'{S}ankar\=abhara\d{n}am}}
\newcommand{\BhairavI}{{\it Bhairav\=i}}

\newcommand{\TODI}{{\it T\=o\d{d}\=i}}

\newcommand{\KAmbhOji}{{\it K\=ambh\=oji}}

\newcommand{\KalyANI}{{\it Kaly\=a\d{n}\=i}}

\newcommand{\SahAnA}{{\it Sah\=an\=a}}
\newcommand{\tabref}[1]{Table \ref{#1}}
\newcommand{\figref}[1]{Fig. \ref{#1}}
\newcommand{\eqnref}[1]{(\ref{#1})}

\name{Venkata Subramanian Viraraghavan,$^{1,2}$
      Arpan Pal,$^{1}$
      R Aravind,$^{2}$ 
      Hema Murthy$^{3}\sthanks{This research was partly funded by the European Research Council under the European Unions Seventh Framework Program, as part of the CompMusic project (ERC grant agreement 267583).}$}
\address{$^1$ TCS Research and Innovation, Embedded Systems and Robotics, Bangalore, India\\
venkatasubramanian.v@tcs.com, arpan.pal@tcs.com\\
         $^2$ Department of Electrical Engineering, $^3$ Department of Computer Science and Engineering\\ Indian Institute of Technology, Madras\\
         aravind@ee.iitm.ac.in, hema@cse.iitm.ac.in
}

\begin{document}

\ninept
\maketitle

\begin{sloppy}

\begin{abstract}
  \Gamaka s are an integral aspect of Carnatic Music, a form of classical music prevalent in South India. They are used in \rAga s, which may be seen as melodic scales and/or a set of characteristic melodic phrases. \Gamaka s exhibit continuous pitch variation often spanning several semitones.

In this paper, we study how \gamaka s scale with tempo and propose a novel approach to change the tempo of Carnatic music pieces. The music signal is viewed as consisting of constant-pitch segments and transients. The transients show continuous pitch variation and we consider their analyses from a theoretical stand-point. We next observe the non-uniform ratios of time-scaling of constant-pitch segments, transients and silence in excerpts from nine concert renditions of \varNa s in six \rAga s.

The results indicate that the changing tempo of Carnatic music does not change the duration of transients significantly. We report listening tests on our algorithm to slow down Carnatic music that is consistent with this observation.

\end{abstract}

\begin{keywords}
Carnatic Music, Pitch transients, Time-scaling.
\end{keywords}

\section{Introduction}
\label{sec:introduction}
\Gamaka s form an integral part of Indian classical music. These are continuous pitch variations that traverse pitches between notes in typical musical scales. In Carnatic music, \gamaka s carry important information relating to the definition and identity of a \rAga~ (a \rAga~ is roughly comparable to scales in Western classical music). It is important that \gamaka s are rendered accurately to preserve the nuances of a raga.

Descriptions of \gamaka s in Carnatic musicology texts (\cite{ssp} is believed to be the first, but \cite{cdp} names at least one earlier source) give a feel for what they are. For example, the \kampita~ \gamaka~ (regarded as the most representative one in Carnatic music) is described in \cite{ssp} as \textit{``Keeping the fingers of the left hand on any svara sth\={a}na [fret] in the v\={i}\d{n}a [a fretted Carnatic instrument] with the m\={i}\d{t}\d{t}u [pluck] and shaking the string is kampita.''}. The `shake' can span over three semitones. However, these descriptions are not directly useful in a mathematical characterization, as was recognized in the CompMusic project \cite{compMusicProject}.

Although Carnatic music is considered replete with \gamaka s, even seasoned practitioners agree that some \rAga s are `\gamaka-heavy' while others are not. Thus, some interesting questions arise:
\begin{enumerate}
\item How much of Carnatic music consists of \gamaka s?
\item How do \gamaka s scale with tempo?
\item Do \gamaka s influence emotional responses to music?
\end{enumerate}

We show in this paper that even for \rAga s seen as \gamaka-heavy, the time-scaling is not uniform. As a result, the first question cannot be answered without a tempo being assumed. The third question is related to the first and we come back to it in Section \ref{sec:conclusion}. Our analysis would conceptually be applicable to any genre of music with \gamaka-equivalent features, notably Hindustani music.


An important compositional form of Carnatic music, the \varNa~ \cite{krishnaRajVarnamsNcc}, is usually sung in two speeds (at least roughly the first one-third is) and are thus ideal for our analysis to answer the second question above. The only notation from Carnatic music we will use in this paper is given in \tabref{tab:svaraNames}.

\begin{table}
 \begin{center}
 \caption{\Svara~ names and positions in the 12 notes of an octave for Carnatic and Western music. The tonic is assumed to be C.}
 \label{tab:svaraNames}
\begin{tiny}
\setlength\tabcolsep{3pt} 
 \begin{tabular}{|c|c|c|c|c|c|c|c|c|c|c|c|c|}
  \hline
  Name & Sa & \multicolumn{2}{|c|}{Ri}  & \multicolumn{2}{|c|}{Ga} & \multicolumn{2}{|c|}{Ma} & Pa & \multicolumn{2}{|c|}{Da} & \multicolumn{2}{|c|}{Ni} \\
  \hline
  Carnatic & S &   R1  & R2 & G2 & G3 & M1 & M2 & P & D1 & D2 & N2 & N3 \\
  \hline
  Western & C &   C\# & D & D\# & E & F & F\# & G & G\# & A & A\# & B \\
  \hline
 \end{tabular}
\end{tiny}
\end{center}
\end{table}
The rest of the paper is organized as follows. Section \ref{sec:previousWork} describes relevant previous work, while windowing-based techniques to track \gamaka s are analyzed in Section \ref{sec:transientAnalysis}. Additional results from analysis of concert recordings in Section \ref{sec:nonUniform} to support an alternative approach to time-scaling in Carnatic music. The results of listening tests to evaluate this technique, presented in Section \ref{sec:results}, are followed by a discussion in Section \ref{sec:conclusion}.

\section{Previous work}
\label{sec:previousWork}
If there is continuous pitch variation, as in \gamaka s, a natural approach is to track the curve(s) the variation follows. Piece-wise linear fitting was used in \cite{vviraraghavanfrsm2004} for retrieval, while Bezier curves were used to characterize them for synthesis \cite{batteyBezier}. In \cite{kGanguliNcc2017}, the variation was `quantized' to eight cubic polynomial curves for retrieval. However, none of them characterizes \gamaka s satisfactorily

There have also been many studies of Western classical music vibratos. An extensive survey of results relating to vibratos can be found in \cite{sundberg1995acoustic}.
However, though similar in spirit to \gamaka s, vibratos are much faster and the range of pitch variation is much smaller and techniques developed for analysis of vibrato do not scale for \gamaka~ analysis. A recent article \cite{hope2016possibilities} describes glissandos, but they are really counterparts to \jAru s, which is a particular type of \gamaka~ in Carnatic music. The most representative \gamaka, \kampita, while prolific in Carnatic music \cite{krishna2012carnatic}, is not used much in Western classical music.

Apart from \gamaka~ analysis, tempo-change related work is relevant. Subramian et. al. \cite{subramanian2011modeling} aim to automatically double the speed of one \varNa. This will be studied in greater detail in Section \ref{sec:nonUniform}. Alternatively, there are software packages that change the tempo of a musical piece from an existing audio sample rather than from a score. Transcribe \cite{transcribe} is one such and Audacity \cite{audacity}, another. While the outputs may seem acceptable for Western classical music, they are inadequate for Carnatic music. 
We will refer work related to musical-emotion when needed in Section \ref{sec:conclusion}.



\section{Transient Analysis}
\label{sec:transientAnalysis}
A case of continuous-pitch variation of a single tone is analyzed in this section. A single tone suffices if it is assumed that harmonics follow the same movement simultaneously, which is observed in \gamaka s. Let it start at a frequency $f_0$ Hz and remain there for $t_{c1}$ seconds. Then, let it make a transition from $f_0$ to $f_1$ and back to $f_0$ in $t_T$ seconds and remain at $f_0$ for $t_{c2}$ seconds. Typical values of all variables are shown in \tabref{tab:typicalValues}; this is for a \kampita~ \gamaka~ by a male voice, such as the Ni in \bhairavI.


\begin{table}
 \begin{center}
 \caption{Typical variable values of lower Ni in \bhairavI~ for a male voice. See \figref{fig:gamakaWindows} for the meanings of the variables.}
 \label{tab:typicalValues}
  \begin{tabular}{|p{1cm}|l|l|l|l|l|}
  \hline
  Variable & $f_0$  &   $f_1$  & $t_{c1}$ & $t_{c2}$ &   $t_T$ \\
  \hline
  Value & 125 Hz & 150 Hz &  $\ge$ 70 ms &  $\ge$ 70 ms & 200 ms \\
  \hline
 \end{tabular}
\end{center}

\end{table}

\begin{figure}
 \centerline{\framebox{
 \includegraphics[width=\columnwidth]{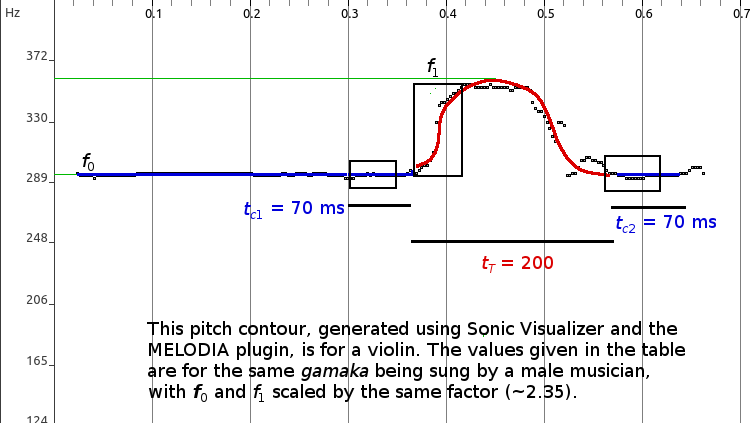}}}
 \caption{Pitch contour for a phrase DN in \bhairavI. Three positions of the analysis window, of width $W = 40$ ms, are shown. The x-axis is time in seconds and the y-axis, frequency in Hz.}
 \label{fig:gamakaWindows}
\end{figure}

The window function $h(t)$ (its Fourier transform as a function of angular frequency $\Omega$ is denoted by $H(\Omega)$), the window size, $W$, i.e. $h(t)$ exists in $(0,W]$, and the window shift, $w$ constitute the remaining parameters of analysis. The typical values for $W$ is $100$ ms \cite{songSegCMMR}. For window shifts of about $w = 100$ ms, the transient will get sampled at most thrice and there is little meaning in trying to trace its \textit{curve}\footnote{Unless we make assumptions about the curve, which we do not.}. In fact, we show that there is little meaning to trace it for any shift size. We do this by examining three positions of the window ($W=40$ ms) as shown in \figref{fig:gamakaWindows}.

In the left-most position of the window, it is safe to assume (because the pitch is not changing) that the time-windowed, digital signal can be viewed as:
\begin{equation} \label{eqn:timeDomainWinLeft}
s_1[n] = h[n] \times a_1 \cos [2 \pi f_0  n T_s  + \theta_1]
\end{equation}
where $a_1$ is the constant amplitude of the signal, $\theta_1$ is a random phase parameter, and $T_s$ is the uniform sampling period.

We call such segments constant-pitch segments or CP-notes. A working definition is \textit{the longest sequence of pitch values whose minimum and maximum are within $0.3$ semitones of the mean, while the magnitude of the slope of the best fit line through those values does not exceed 1 semitone per second}. The short-term Fourier transform of $s_1[n]$ is:

\newcommand{\ejw}{e^{j\omega}}

\begin{equation} \label{eqn:freqDomainWinRight}
S_1(\ejw) = a_1 \frac{H(e^{j(\omega - 2 \pi f_0T_s)})e^{j \omega \theta_1} + H(e^{j(\omega + 2 \pi f_0T_s)})e^{-j\omega \theta_1}}{2}
\end{equation}

Similar equations would apply for the right-most position of the window, except that $\theta_2 \neq \theta_1$ and $a_1 \neq a_2$ in general.

The form of \eqnref{eqn:freqDomainWinRight} is amenable to reasonably accurate pitch-tracking, provided the main-lobe width of $H(\ejw)$, $B_H$, satisfies $B_H \le 2 \pi f_0 T_s$. This last condition can be assured by choosing a long-enough window size.

For the `middle' position of the window, an equation such as \eqnref{eqn:timeDomainWinLeft} is much harder to formulate. However, we do know that the signal within the window has a variation in frequency, expressed as a ratio $\rho$, which must equal or exceed in parts that of the linear variation from $f_0$ to $f_1$ (in duration $t_T/2$). That is, $\rho \ge \rho_L$ where
\begin{equation}\label{eqn:ratioInWindow}
\rho_L = 1 + \frac{f_1 - f_0}{f_0}\times\frac{W}{t_T/2} = 1 + \frac{1}{15} = 1.0\bar{6}
\end{equation}

This ratio seems small at first sight, but it is greater than a semitone. In fact, the pitch in this window cannot be \textit{defined} because its spectrum would have energies spread between $f_0$ and $f_0(1+\rho)$. Further, it is impractical to progressively decrease the window size to reduce $\rho$ because the resolution of nearby frequencies (i.e. harmonics in case of real musical signals) worsens, which is due to widening of the main lobe in the window's spectrum. Only the very slow \gamaka s (like a slow \jAru, or `slide') can be analyzed in this manner. It does not work for the typical range of transient durations ($100$ to $200$ ms).


This analysis suggests that \textit{transients should not be analyzed by reducing the time-shift between windows ($w$) while leaving ($W$) large}. Fortunately, in Carnatic music, there may not be the need to \textit{precisely find} the curve of a \gamaka, as we found in informal experiments, and is supported by the number of successful interpolation techniques for synthesis \cite{ishwar2013motif, batteyBezier, subramanian2011modeling}.


\section{Non-uniform scaling of Transients and CP-Notes}
\label{sec:nonUniform}
The foregoing analysis leads to an interesting question: Do transients slow down with tempo? Surprisingly, it appears that they scale to only within a limited range as we show below.

\subsection{A close look at time-scaling of transients}
Consider the two speeds of rendering a single \svara, Ri, in \kEdAragauLa~ \rAga. The pitch contours (estimated from \cite{audacity}) are shown in \figref{fig:kGaulaiRiNotAligned}. Transients are manually marked with red curves and CP-notes, with blue lines. It is fairly evident from the figure that the transients and CP-notes do not scale the same way. To aid seeing this better, \figref{fig:kGaulaiRiAligned} has the pitch contours of the faster renditions shifted so that the transients are aligned. The figure clearly shows that CP-notes are scaled down much more than the transients. This is analogous to speech, where the duration of consonants is preserved across different speeds, while the duration of vowels suffers significantly. In order to identify words that are spoken, consonants are necessary.  The importance of vowels cannot be undermined, though: the distinction between beet and bit is just the vowel.


\subsection{Non-uniform scaling in longer examples}
The examples discussed so far were analyzed manually.   The scalability of this approach is established by analyzing varnams in six \rAga s.   The pitch contour of a given \varNa~  is  first estimated using the MELODIA algorithm \cite{melodia}. The pitch contour is further segmented into CP-notes, silence (corresponds to regions where the pitch estimate is zero) and transients (everything other than CP-notes and silence).   The \varNa s are a particular class of items, where some lines (about a third of the composition) are usally rendered in at least at two different speeds.   The durations of CP-segments, transients and silence segments in the first and second speeds are obtained.    

Next, the ratios of the durations in the first speed to the corresponding durations in the second speed were found. The overall ratio is defined as the total duration in the first speed to that in the second\footnote{The second speed is actually twice that of the first, but it is common practice that lines of the \varNa~ are repeated in the first speed and not in the second. This explains why the overall ratio is well above $2$ in \tabref{tab:cpRatiosAllVarnas}.}. \tabref{tab:cpRatiosAllVarnas} shows very clearly that there is a large difference in the ratios of the CP-notes and transients (and silence).

\subsection{Proposed Algorithm} \label{sec:proposedAlgo}
Based on the observations presented, we propose Algorithm \ref{algo:slowDown} to slow down music by a factor $R \ge 1$. The actual slowing down (statement \ref{stmt:psola}) is similar to the TD-PSOLA algorithm \cite{moulines1990pitch, huang2001spoken}. However, that is not the main contribution of the algorithm. Recalling the analysis in Section \ref{sec:transientAnalysis}, it is to be noted that when dealing with transients, it is often not possible to define the pitch curve exactly. This immediately implies that a pitch-synchronous method of time-scaling transients would be ambiguous. Instead, as a first approximation, we \textit{do not scale transients at all} (statement \ref{stmt:noScaling}). The algorithm is given for the case of integer $R$, but it can be appropriately extended to other cases. The tonic, $f_0$, was found manually.

Algorithm \ref{algo:slowDown} was used after `snapping' stationary points to the nearest CP-notes, which cluster around scale notes (\tabref{tab:svaraNames}) with very sharp peaks \cite{viraraghavan2017Ismir}. If the pitch values for $80$ ms around a stationary point were within $0.3$ semitones from any of the CP-note peaks, they are counted as a CP-note (the slope condition in the definition of a CP-note is not enforced). Second, based on the the result in \cite{swathithesis}, CP-notes shorter than $250$ milliseconds are not extended beyond this limit. This technique results in an effective slowing-down factor being $R' < R$. In the experiments described in Section \ref{sec:results}, $R'$ ranged from $1.79$ to $1.81$ for $R=2$.

Clearly, Algorithm \ref{algo:slowDown} is only one way of slowing down music, but the result in Section \ref{sec:results} shows equally clearly that uniform scaling of transients is not the way to do it.
Non-uniform scaling of the transients within the limits of \tabref{tab:cpRatiosAllVarnas} may help for a larger range of $R$, but it is beyond the scope of this paper.

\begin{algorithm}
\begin{algorithmic}[1]
\STATE Segment the music samples into non-overlapping frames of length $W = 32$ ms.
\STATE Find the, say $K$, silence segments \cite{vviraraghavanfrsm2004}, each lasting from $x_k$ to $y_k$ frames, $1 \le k \le K$.
\STATE Track the pitch in each frame \cite{vviraraghavanfrsm2004} (algorithm modified to use phase-information of the spectrum) to obtain $f[l], 0 \le l < L$.
\STATE In the regions of music (i.e. not silence), find the pitch in semitones with respect to the tonic, $f_0$, as $n[l] = 12\log_2(\frac{f[l]}{f_0})$.
\STATE Identify, say $C$, CP-notes according to the definition in Section \ref{sec:transientAnalysis}. Let the $j^{\mbox{th}}$ CP-note start at frame $c_j$ and end at frame $d_j$.
\FOR{$1 \le j \le C$}
\STATE Mark the nominal start and end of the $j^{\mbox{th}}$ slowed-down CP-note as $\hat{c}_j \leftarrow Rc_j$ and $\hat{d}_j \leftarrow Rd_j$.
\ENDFOR 
\STATE Identify, say $I$, transients (other than silence and CP-notes). Let each start at frame $s_i$ and end at frame $e_i$.
\FOR{$1 \le i \le I$}
\STATE Set the beginning and end of the slowed-down transient as $\hat{s}_i \leftarrow Rs_i$ and $\hat{e}_i \leftarrow (R-1)s_i + e_i$.
\label{stmt:noScaling}
\STATE Find the nearest CP-note or silence segment on either side of the $i^{\mbox{th}}$ transient.
\IF{an earlier CP-note (indexed by $j(i)$) or silence (indexed by $k(i)$) flanks the transient}
\STATE $\hat{d}_{j(i)} \leftarrow \hat{s}_i - 1$ or $\hat{y}_{j(i)} \leftarrow \hat{s}_i - 1$.
\ENDIF
\IF{a later CP-note (indexed by $j(i)$) or silence (indexed by $k(i)$) flanks the transient}
\STATE $\hat{c}_{j(i)} \leftarrow \hat{e}_i - 1$ or $\hat{y}_{j(i)} \leftarrow \hat{e}_i - 1$.
\ENDIF
\ENDFOR
\FOR{$1 \le j \le C$}
\STATE Divide the signal by the interpolated amplitude of the CP-note (energy in frames $c_j$ to $d_j$).
\STATE Interpolate the original amplitude from frames $\hat{c}_j$ to $\hat{d}_j$.
\STATE Find the numbers of frames of attack and decay of the CP-note. Let these be $a_j (\ge 1)$ and $b_j (\ge 1)$ respectively.
\STATE Pitch-synchronously, extend the steady part of the CP-note in frames $c_j + a_j$  to $d_j - b_j$ to occupy frames $\hat{c}_j + a_j$ to $\hat{d}_i - b_j$.
\label{stmt:psola}
\STATE Multiply the signal by the interpolated amplitude.
\STATE Copy the signal from frames $c_j$ to $c_j + a_j - 1$ to frames $\hat{c}_j$ to $\hat{c}_j + a_j - 1$ and similarly for frames $d_j - b_j + 1$ to $b_j$.
\ENDFOR
\FOR{$1 \le k \le K$}
\STATE Extend the signal in frames $x_k + 1$ to $y_k - 1$ by repetition (with any excess repetition deleted) to occupy frames $\hat{x}_k+1$ to $\hat{y}_k - 1$.
\STATE Copy the signal from frames $x_k$ and $y_k$ to frames $\hat{x}_k$ and $\hat{y}_k$ respectively.
\ENDFOR
\end{algorithmic}
\caption{Non-uniform slowing down of Carnatic music}
\label{algo:slowDown}
\end{algorithm}


\begin{figure*}
  \centering
  \begin{subfigure}[b]{\columnwidth}
  \centerline{\includegraphics[width=\columnwidth]{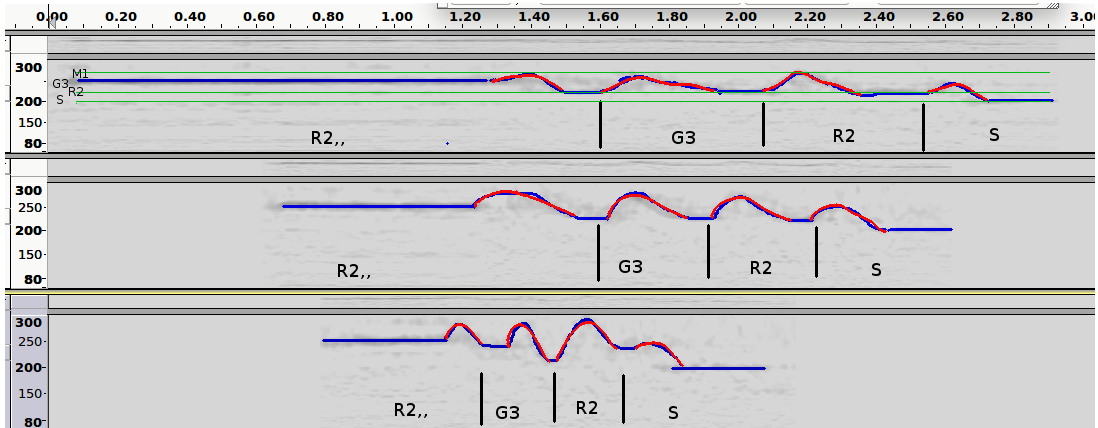}}
  \caption{Approximately aligned at the centers of the phrase in each speed.}
  \label{fig:kGaulaiRiNotAligned}
  \end{subfigure}
  ~
  \begin{subfigure}[b]{\columnwidth}
  \centerline{\includegraphics[width=\columnwidth]{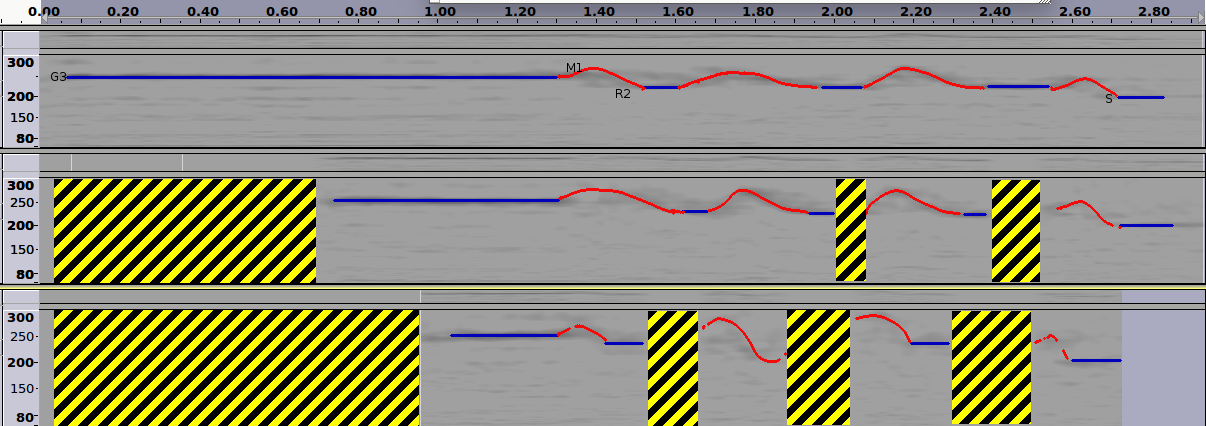}}
  \caption{Silence (hatching) introduced for alignment at transients.}
  \label{fig:kGaulaiRiAligned}
  \end{subfigure}
  \caption{Pitch contours of the rendering of the phrase RGRS in the \rAga~ \kEdAragauLa, in three speeds. Transients have been manually marked with red curves, and CP-notes, with blue lines. The x-axis is time in mm:ss format and the y-axis, frequency in Hz.}
  \label{fig:kGaulaiRi}
\end{figure*}


\begin{table*}
\centering
\caption{Durations and ratios of constant-pitch segments and transients in two speeds of the first parts of several \varNa s in six \rAga s.}
\label{tab:cpRatiosAllVarnas}
\begin{tabular}{|p{2.5cm}|p{1.2cm}||p{1.2cm}|p{1.2cm}|p{1.2cm}|p{1.2cm}||p{1.2cm}|p{1.2cm}|p{1.2cm}|p{1.0cm}|}
\hline
\RAga & Number of \varNa s & \multicolumn{4}{|c||}{Duration in the $1^{\mbox{st}}$ speed (seconds)} & \multicolumn{4}{|c|}{Ratio ($1^{\mbox{st}}$ speed to $2^{\mbox{nd}}$ speed)} \\
\cline{3-10}
 &  & CP-notes & Transients & Silence & Overall & CP-notes & Transients & Silence & Overall \\ \hline
\TODI & 1 & 37.7 & 69.7 & 10.5 & 117.9 & 8.66 & 2.61 & 3.79 & 3.49 \\ \hline
\BhairavI & 1 & 56.0 & 100.2 & 64.1 & 220.3 & 10.27 & 2.58 & 5.44 & 3.93 \\ \hline
\KAmbhOji &1 & 52.1 & 79.3 & 29.1 & 160.5 & 6.2& 1.9& 2.8& 2.7\\ \hline
\ZankarAbharaNa & 1 & 43.6 & 87.1 & 11.7 & 142.4 & 12.3 & 3.48 & 2.33 & 4.23 \\ \hline
\SahAnA &3 & 144.4 & 237.2 & 78.9 & 460.5 & 9.5 & 2.8 & 3.8 & 3.8\\ \hline
\KalyANI & 2 & 114.6 & 220.1 & 59.6 & 394.3 & 10.1 & 3.0 & 4.5 & 4.1 \\ \hline
\end{tabular}
\end{table*}

\section{Experiments and Results}
\label{sec:results}
The algorithm described in Section \ref{sec:proposedAlgo} (with the stated modifications) was implemented on $1.5$ min-long, de-noised (using \cite{audacity}) audio samples in three \rAga s \cite{ragasurabhiBhairavi, ragasurabhiSahana, ragasurabhiShankarabharanam}. Each output was split into two $1$-minute clips. Similarly, the outputs from an existing, uniform slowing-down algorithm in \cite{audacity} were also split at the same locations. The slowing-down factor given to the existing algorithm was $R'$ (not $R$), which could vary piece by piece.

The original clip was also made into two clips according to the split in the slowed-down pieces. The resulting $18$ clips were played in a blind listening test\footnote{\url{https://www.iitm.ac.in/donlab/pctestmusic/index.html?owner=venkat&testid=test1&testcount=6}} where participants were asked to rank the slowed-down clips on a scale of 1 (worst) to 5 (best) relative to the original clip. The order of the slowed-down clips was random. Participants also rated their own familiarity with the \rAga s.

The result is strongly in favor of non-uniform scaling. Eighteen users ($12$ experts) took the test and the proposed algorithm was preferred to the existing one in $84\%$ of the cases ($90\%$ among experts). The average rating of the proposed algorithm was $3.6$ (experts: $3.74$) and that of the existing one was $2.45$ (experts: $2.38$). The small difference in preference suggests that the experts based their evaluations on \rAga-identity more than others did.

\section{Discussion}
\label{sec:conclusion}
We conclude with a discussion touching on the three questions mentioned in Section \ref{sec:introduction}. It is clear that transients scale non-uniformly with tempo and that immediately implies that the fraction of music with \gamaka s (as against CP-notes) would vary. For any conclusion more specific than `the fraction of music containing transients increases with speed,' future studies are needed.

In \cite{subramanian2011modeling}, which considered speed doubling, the synthesis step (from notation) was preceded by a manual analysis of \gamaka s. The first example we considered in Section \ref{sec:nonUniform} shows that it is useful to view a \svara~ with a \gamaka, as consisting of a CP-note and one or more transients. In the language of \cite{subramanian2011modeling}, we believe that focal pitches of CP-note segments of a \gamaka~ must be treated differently from focal pitches in transients. 

We proposed an algorithm that does not scale transients when slowing down Carnatic music. This was clearly preferred by listeners over uniform slowing down. For future work, we propose that tempo change algorithms should be parameterized with a \textit{priority choice} between CP-notes (which we prioritized in our algorithm) and transients. This may depend on the \rAga: e.g, \bhairavI~ and \mOhana~ could prioritize transients  and CP-notes respectively. Thus, it is not always a case of changing `detail' as noted in \cite{subramanian2011modeling}.


Finally, previous work on musical emotion in Carnatic and Hindustani music \cite{mathur2015emotional, koduri2010behavioral} has worked at the level of \rAga s. Yet, for example, in \tODI, long stretches of \svara s that eschew S and P build up `tension' and a transient-heavy \gamaka~ can release it to give a calming effect.  A future, underlying mechanism of musical emotion in Indian music (see  \cite{juslin2008emotional} for its need in Western music) should thus account for the effects of CP-notes and transients.

\section{Acknowledgments}
The authors thank the Raga Surabhi team (\url{http://www.ragasurabhi.com/index.html}) for their kind permission to use three recordings \cite{ragasurabhiBhairavi, ragasurabhiSahana, ragasurabhiShankarabharanam} for the experiments described in this paper, and all participants of the listening test. V Viraraghavan thanks Ms. Anju Leela Thomas for her help in setting up this test.

\bibliographystyle{IEEEtran}
\bibliography{venkat.bib}
%
%
%
%
%
%
%
%
%

\end{sloppy}
\end{document}